\documentclass[twocolumn,twoside,slac_two]{revtex4}
\usepackage{graphicx}
\usepackage{fancyhdr}
\begin{document}



\def\Journal#1#2#3#4{{#1} {\bf #2} (#4) #3}

\def\NCA{\em Nuovo Cimento}
\def\NIM{\em Nucl.~Instrum.~Methods}
\def\NIMA{{\em Nucl.~Instrum.~Methods}~A}
\def\NIMB{{\em Nucl.~Instrum.~Methods}~B}
\def\NPB{{\em Nucl.~Phys.}~B}
\def\PLB{{\em Phys.~Lett.}~B}
\def\PRL{\em Phys.~Rev.~Lett.}
\def\PRD{{\em Phys.~Rev.}~D}
\def\ZPC{{\em Z.~Phys.}~C}
\def\EPJ{{\em Eur.~Phys.~J.}~C}
\def\JPG{{\em Jour.~of~Phys.}~G}
\def\JHEP{{\em JHEP}}
\def\PoS{{PoS}}

\newcommand{\op}[1]{\mbox{\bf #1}}
\newcommand{\bra}{\langle}
\newcommand{\ket}{\rangle}

\newcommand{\subrm}[1]{\mbox{\tiny \rm #1}}
\newcommand{\CP}{{\cal CP}}
\newcommand{\Vus}{{|V_{us}|}}
\newcommand{\um}{\mu{\rm m}}
\newcommand{\Br}{{\text{Br}}}
\newcommand{\bfBr}{{\text{\bf \small Br}}}
\newcommand{\epe}{\epsilon'/\epsilon\,}
\newcommand{\Reepe}{\text{Re(}\epsilon'/\epsilon \rm{)}}
\newcommand{\reeta}{\text{Re(}\eta_{000}\rm{)}}
\newcommand{\relli}{R_{\subrm{ellipse}}}
\newcommand{\Relli}{\relli}
\newcommand{\imeta}{\rm{Im(}\eta_{000}\rm{)}}
\newcommand{\alphak}{\alpha_{K^*}}
\newcommand{\pid}{\pi^0_{\subrm{Dalitz}}}
\newcommand{\KKbar}{\overline{K^0} K^0}
\newcommand{\Ml}{M_{\subrm{L}}}
\newcommand{\Ms}{M_{\subrm{S}}}
\newcommand{\Mls}{M_{\subrm{L,S}}}
\newcommand{\Msl}{M_{\subrm{S,L}}}
\newcommand{\ml}{m_{\subrm{L}}}
\newcommand{\ms}{m_{\subrm{S}}}
\newcommand{\mls}{m_{\subrm{L,S}}}
\newcommand{\msl}{m_{\subrm{S,L}}}
\newcommand{\gammal}{\Gamma_{\subrm{L}}}
\newcommand{\gammas}{\Gamma_{\subrm{S}}}
\newcommand{\gammals}{\Gamma_{\subrm{L,S}}}
\newcommand{\gammasl}{\Gamma_{\subrm{S,L}}}
\newcommand{\kl}{K_{\subrm{L}}}
\newcommand{\ks}{K_{\subrm{S}}}
\newcommand{\kls}{K_{\subrm{L,S}}}
\newcommand{\ksl}{K_{\subrm{S,L}}}
\newcommand{\kspiee}{\ks \to \pi^0 \, e^+ \, e^-}
\newcommand{\klpiee}{\kl \to \pi^0 \, e^+ \, e^-}
\newcommand{\klspiee}{\kls \to \pi^0 \, e^+ \, e^-}
\newcommand{\kspimumu}{\ks \to \pi^0 \, \mu^+ \, \mu^-}
\newcommand{\klpimumu}{\kl \to \pi^0 \, \mu^+ \, \mu^-}
\newcommand{\klspimumu}{\kls \to \pi^0 \, \mu^+ \, \mu^-}
\newcommand{\kspipipi}{\ks \to 3 \pi^0}
\newcommand{\klpipipi}{\kl \to 3  \pi^0}
\newcommand{\klspipipi}{\kls \to 3  \pi^0}
\newcommand{\kspipipid}{\ks \to \pi^0 \pi^0 \pid}
\newcommand{\klpipipid}{\kl \to \pi^0 \pi^0 \pid}
\newcommand{\klspipipid}{\kls \to \pi^0 \pi^0 \pid}
\newcommand{\kspipipic}{\ks \to \pi^+ \pi^- \pi^0}
\newcommand{\klpipipic}{\kl \to \pi^+ \pi^- \pi^0}
\newcommand{\klspipipic}{\kls \to \pi^+ \pi^- \pi^0}
\newcommand{\klpipi}{\kl \to \pi^+ \pi^-}
\newcommand{\kspipi}{\ks \to \pi^+ \pi^-}
\newcommand{\klspipi}{\kls \to \pi^+ \pi^-}
\newcommand{\klpizpiz}{\kl \to \pi^0 \pi^0}
\newcommand{\kspizpiz}{\ks \to \pi^0 \pi^0}
\newcommand{\klspizpiz}{\kls \to \pi^0 \pi^0}
\newcommand{\kspipid}{\ks \to \pi^0 \pid}
\newcommand{\kspidpid}{\ks \to \pid \pid}
\newcommand{\klpipid}{\kl \to \pi^0 \pid}
\newcommand{\klpidpid}{\kl \to \pid \pid}
\newcommand{\kgg}{K \to \gamma \gamma}
\newcommand{\ksgg}{\ks \to \gamma \gamma}
\newcommand{\klgg}{\kl \to \gamma \gamma}
\newcommand{\klsgg}{\kls \to \gamma \gamma}
\newcommand{\kslgg}{\ksl \to \gamma \gamma}
\newcommand{\klggg}{\kl \to \gamma \gamma \gamma}
\newcommand{\klthreeg}{\kl \to 3 \gamma}
\newcommand{\kspigg}{\ks \to \pi^0 \gamma \gamma}
\newcommand{\klpigg}{\kl \to \pi^0 \gamma \gamma}
\newcommand{\klpipig}{\kl \to \pi^+ \pi^- \gamma}
\newcommand{\kspipig}{\ks \to \pi^+ \pi^- \gamma}
\newcommand{\klpizpizg}{\kl \to \pi^0 \pi^0 \gamma}
\newcommand{\kspipiee}{\ks \to \pi^+ \pi^- e^+ e^-}
\newcommand{\klpipiee}{\kl \to \pi^+ \pi^- e^+ e^-}
\newcommand{\klspipiee}{\kls \to \pi^+ \pi^- e^+ e^-}
\newcommand{\klspienu}{\kls \to \pi^\pm e^\mp \nu_e(\bar{\nu_e})}
\newcommand{\klpienu}{\kl \to \pi^\pm e^\mp \nu_e(\bar{\nu_e})}
\newcommand{\kspienu}{\ks \to \pi^\pm e^\mp \nu_e(\bar{\nu_e})}
\newcommand{\kpienu}{K^0 \to \pi^+ e^- \nu_e}
\newcommand{\kbarpienu}{\overline{K^0} \to \pi^- e^+ \bar{\nu_e}}
\newcommand{\klpipie}{\kl \to \pi^\pm \pi^0 e^\mp \nu_e(\bar{\nu_e})}
\newcommand{\klpipimu}{\kl \to \pi^\pm \pi^0 \mu^\mp \nu_{\mu}(\bar{\nu_{\mu}})}
\newcommand{\kleeee}{\kl \to e^+ e^- e^+ e^-}
\newcommand{\klfoure}{\kl \to 4 e}
\newcommand{\klmumuee}{\kl \to \mu^+ \mu^- e^+ e^-}
\newcommand{\klee}{\kl \to e^+ e^-}
\newcommand{\klmumu}{\kl \to \mu^+ \mu^-}
\newcommand{\kleeg}{\kl \to e^+ e^- \gamma}
\newcommand{\kleegg}{\kl \to e^+ e^- \gamma \gamma}
\newcommand{\klmumug}{\kl \to \mu^+ \mu^- \gamma}
\newcommand{\klpinunu}{\kl \to \pi^0 \nu \bar{\nu}}
\newcommand{\kpinunu}{K \to \pi \nu \bar{\nu}}
\newcommand{\kpmmunu}{K^\pm \to \mu^\pm \nu_\mu}
\newcommand{\kpmunu}{K^+ \to \mu^+ \nu_\mu}
\newcommand{\kmmunu}{K^- \to \mu^- \bar{\nu}_\mu}
\newcommand{\kpmenu}{K^\pm \to e^\pm \nu_e}
\newcommand{\kpenu}{K^+ \to e^+ \nu_e}
\newcommand{\kmenu}{K^- \to e^- \bar{\nu}_e}
\newcommand{\Remu}{K_{e2}/K_{\mu2}}
\newcommand{\Gemu}{\Gamma(K_{e2})/\Gamma(K_{\mu2})}
\newcommand{\Rmue}{K_{\mu3}/K_{e3}}
\newcommand{\Gmue}{\Gamma(K_{\mu3})/\Gamma(K_{e3})}
\newcommand{\kpmpipipi}{K^\pm \to \pi^\pm \pi^+ \pi^-}
\newcommand{\kppipipi}{K^+ \to \pi^+ \pi^+ \pi^-}
\newcommand{\kmpipipi}{K^- \to \pi^- \pi^- \pi^+}
\newcommand{\kpmpipizpiz}{K^\pm \to \pi^\pm \pi^0 \pi^0}
\newcommand{\kppipizpiz}{K^+ \to \pi^+ \pi^0 \pi^0}
\newcommand{\kmpipizpiz}{K^- \to \pi^- \pi^0 \pi^0}
\newcommand{\kpmpipiz}{K^\pm \to \pi^\pm \pi^0}
\newcommand{\kppipiz}{K^+ \to \pi^+ \pi^0}
\newcommand{\kmpipiz}{K^- \to \pi^- \pi^0}
\newcommand{\kpmpienu}{K^\pm \to \pi^0 e^\pm \nu (\bar{\nu})}
\newcommand{\kppienu}{K^+ \to \pi^0 e^+ \nu}
\newcommand{\kmpienu}{K^- \to \pi^0 e^- \bar{\nu}}
\newcommand{\kpmpimunu}{K^\pm \to \pi^0 \mu^\pm \nu (\bar{\nu})}
\newcommand{\kppimunu}{K^+ \to \pi^0 \mu^+ \nu}
\newcommand{\kmpimunu}{K^- \to \pi^0 \mu^- \bar{\nu}}
\newcommand{\kpmpipie}{K^\pm \to \pi^+ \pi^- e^\pm \nu (\bar{\nu})}
\newcommand{\kppipie}{K^+ \to \pi^+ \pi^- e^+ \nu}
\newcommand{\kmpipie}{K^- \to \pi^- \pi^+ e^- \bar{\nu}}
\newcommand{\kpmpizpize}{K^\pm \to \pi^0 \pi^0 e^\pm \nu (\bar{\nu})}
\newcommand{\kppizpize}{K^+ \to \pi^0 \pi^0 e^+ \nu}
\newcommand{\kmpizpize}{K^- \to \pi^0 \pi^0 e^- \bar{\nu}}
\newcommand{\kpmpipimu}{K^\pm \to \pi^+ \pi^- \mu^\pm \nu (\bar{\nu})}
\newcommand{\kppipimu}{K^+ \to \pi^+ \pi^- \mu^+ \nu}
\newcommand{\kmpipimu}{K^- \to \pi^- \pi^+ \mu^- \bar{\nu}}
\newcommand{\kpmpizpizmu}{K^\pm \to \pi^0 \pi^0 \mu^\pm \nu (\bar{\nu})}
\newcommand{\kppizpizmu}{K^+ \to \pi^0 \pi^0 \mu^+ \nu}
\newcommand{\kmpizpizmu}{K^- \to \pi^0 \pi^0 \mu^- \bar{\nu}}
\newcommand{\kpmpinunu}{K^\pm \to \pi^\pm \nu \bar{\nu}}
\newcommand{\kppinunu}{K^+ \to \pi^+ \nu \bar{\nu}}
\newcommand{\kmpinunu}{K^- \to \pi^- \nu \bar{\nu}}
\newcommand{\kpmpill}{K^\pm \to \pi^\pm l^+ l^-}
\newcommand{\kppill}{K^+ \to \pi^+ l^+ l^-}
\newcommand{\kmpill}{K^- \to \pi^- l^+ l^-}
\newcommand{\kpmpiee}{K^\pm \to \pi^\pm e^+ e^-}
\newcommand{\kppiee}{K^+ \to \pi^+ e^+ e^-}
\newcommand{\kmpiee}{K^- \to \pi^- e^+ e^-}
\newcommand{\kpmpimumu}{K^\pm \to \pi^\pm \mu^+ \mu^-}
\newcommand{\kppimumu}{K^+ \to \pi^+ \mu^+ \mu^-}
\newcommand{\kmpimumu}{K^- \to \pi^- \mu^+ \mu^-}
\newcommand{\kpmpipiee}{K^\pm \to \pi^\pm \pi^0 e^+ e^-}
\newcommand{\kppipiee}{K^+ \to \pi^+ \pi^0 e^+ e^-}
\newcommand{\kmpipiee}{K^- \to \pi^- \pi^0 e^+ e^-}
\newcommand{\kpmpipig}{K^\pm \to \pi^\pm \pi^0  \gamma}
\newcommand{\kppipig}{K^+ \to \pi^+ \pi^0 \gamma}
\newcommand{\kmpipig}{K^- \to \pi^- \pi^0 \gamma}
\newcommand{\kpmpigg}{K^\pm \to \pi^\pm \gamma  \gamma}
\newcommand{\kppigg}{K^+ \to \pi^+ \gamma \gamma}
\newcommand{\kmpigg}{K^- \to \pi^- \gamma \gamma}
\newcommand{\kpmpipigg}{K^\pm \to \pi^\pm \pi^0 \gamma  \gamma}
\newcommand{\kppipigg}{K^+ \to \pi^+ \pi^0 \gamma \gamma}
\newcommand{\kmpipigg}{K^- \to \pi^- \pi^0 \gamma \gamma}
\newcommand{\kpmpigee}{K^\pm \to \pi^\pm \gamma e^+ e^-}
\newcommand{\kppigee}{K^+ \to \pi^+ \gamma e^+ e^-}
\newcommand{\kmpigee}{K^- \to \pi^- \gamma e^+ e^-}
\newcommand{\xilampi}{\Xi^0 \to \Lambda \pi^0}
\newcommand{\axilampi}{\overline{\Xi^0} \to \overline{\Lambda} \pi^0}
\newcommand{\xilamgam}{\Xi^0 \to \Lambda \gamma}
\newcommand{\axilamgam}{\overline{\Xi^0} \to \overline{\Lambda} \gamma}
\newcommand{\xisiggam}{\Xi^0 \to \Sigma^0 \gamma}
\newcommand{\axisiggam}{\overline{\Xi^0} \to \overline{\Sigma^0} \gamma}
\newcommand{\xisiglnu}{\Xi^0 \to \Sigma^+ l^- \nu}
\newcommand{\axisiglnu}{\overline{\Xi^0} \to \Sigma^- l^+ \bar{\nu}}
\newcommand{\xisigenu}{\Xi^0 \to \Sigma^+ e^- \nu}
\newcommand{\axisigenu}{\overline{\Xi^0} \to \Sigma^- e^+ \bar{\nu}}
\newcommand{\xibeta}{\xisigenu}
\newcommand{\axibeta}{\axisigenu}
\newcommand{\xisigmunu}{\Xi^0 \to \Sigma^+ \mu^- \nu}
\newcommand{\axisigmunu}{\overline{\Xi^0} \to \Sigma^- \mu^+ \bar{\nu}}
\newcommand{\etazz}{\eta_{00}}
\newcommand{\etazzz}{\eta_{000}}
\newcommand{\TeVcc}{TeV/$c^2$}
\newcommand{\GeVcc}{GeV/$c^2$}
\newcommand{\MeVcc}{MeV/$c^2$}
\newcommand{\bdm}{\begin{displaymath}}
\newcommand{\edm}{\end{displaymath}}
\newcommand{\be}{\begin{equation}}
\newcommand{\ee}{\end{equation}}
\newcommand{\pow}[2]{#1 \times 10^{#2}}

\newcommand{\zp}{Z.~Phys.}
\newcommand{\nc}{Nuovo~Cim.}
\newcommand{\np}{Nucl.~Phys.}
\newcommand{\FP}{Fortsch.~der~Physik}
\newcommand{\nw}{Die~Naturwissenschaften}
\newcommand{\nim}{Nucl.~Instr.~and~Meth.}
\newcommand{\spr}{Springer-Verlag, Berlin-Heidelberg-New York}
\newcommand{\spri}{Springer-Verlag, Berlin-Heidelberg-New
York-London-Paris-Tokyo-Hong Kong-Barcelona-Budapest}
\newcommand{\rep}{Phys.~Rep.}

\title{Precision Standard Model Tests with Kaons}

%

\author{R.~Wanke}
\affiliation{Institut f\"ur Physik, Universt\"at Mainz, D-55099 Mainz, Germany}

\begin{abstract}
In kaon physics, several new precision measurements on flavour variables and $\CP$ violation have performed in the recent years.
Presented are a new precise determination of the CKM parameter $\Vus$, which combines the results of all experiments together with recent theoretical progress, and new measurements of the ratio $R_K = \Gamma(K^+ \to e^+ \nu)/\Gamma(K^+ \to \mu^+ \nu)$, which is sensitive to contributions of a possible
charged Higgs in the SUSY framework.
Also, final results of a precision search for direct $\CP$ violation in charged kaon decays are presented.
\end{abstract}

\maketitle

\thispagestyle{fancy}


\section{Introduction}

In the recent years, kaon physics has celebrated a huge revival, 
with many new experiments recording data of several billion kaon decays.
These new data samples allow to study flavour physics and 
$\CP$ violation with even greater precision than in $B$ or $D$ decays, 
as well as to explore many rare and extremely rare decays in
the search for new physics.

In the following, a new precise determination of the CKM parameter $\Vus$, 
new measurements of the ratio $R_K = \Gamma(K^+ \to e^+ \nu)/\Gamma(K^+ \to \mu^+ \nu)$, 
and results of a precision search for direct $\CP$ violation in charged kaon decays are reported.

\section{Precision Measurement of $\Vus$}

The measurement of the parameter $V_{us}$ of the Cabibbo-Kobayashi-Maskawa mixing matrix 
of weak eigenstates has undergone a large improvement with respect to one or two year ago, 
due to both more precise measurements of the main kaon decays
and better computations for the theoretical inputs.

The kaon semileptonic decay width is given by~\cite{bib:Kl3formula}
\begin{eqnarray}
\Gamma(K_{l3(\gamma)}) & \! \! = \! \! & \frac{G_F^2 m_K^5}{192 \pi^3} \, C_K^2 \, S_{EW} \\
                       &   & \times \,  |V_{us}|^2 \, |f_+(0)|^2 I_K^l (1 + 2 \delta^l_{SU(2)} + 2 \delta^l_{EM}),\nonumber
\end{eqnarray}
with $C_K^2 = 1$ ($\frac{1}{2}$) for $K^0$ ($K^\pm$), the short distance electro-weak correction
$S_{EW} = 1.0232$, the hadronic matrix element $f_+(0)$ at $q^2 = 0$, the integral 
$I^l_K$ of the form factors over the phase space, and the form factor corrections
$\delta^l_{SU(2)}$ and $\delta^l_{EM}$ for isospin symmetry breaking and long-distance electro-magnetic interactions, respectively.

The decay width is provided by experiment by measurements of the lifetimes and semileptonic branching fractions
of charged and neutral kaons and
the integrals $I^l_K$ are determined by measurements of the form factor slope parameters.
In particular, the KLOE collaboration has recently provided a complete set of new 
absolute branching fraction measurements~\cite{bib:KLOEbr}, but also the NA48/2 and ISTRA+ experiments
published new precise measurements of the semileptonic $K^\pm$ branching fractions with respect
to the $\pi^\pm \pi^0$ decay~\cite{bib:NA48br,bib:ISTRAbr}.

\begin{figure}[t]
  \begin{center}
    \includegraphics[width=0.7\linewidth]{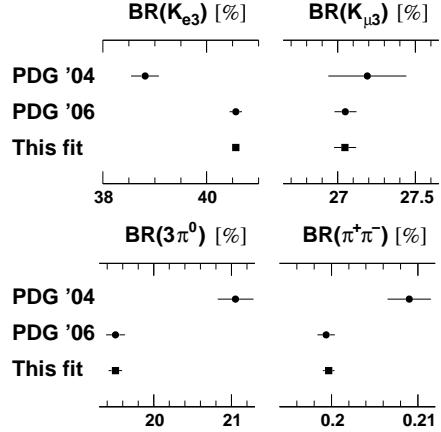}
    \caption{Main $K_L$ branching fractions~\cite{bib:flavianet}.} 
    \label{fig:klbr}
  \end{center}
\end{figure}

\begin{figure}[t]
  \begin{center}
    \includegraphics[width=\linewidth]{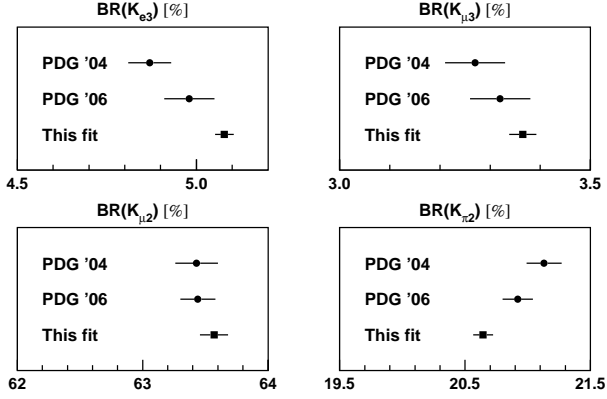}
    \caption{Main $K^\pm$ branching fractions~\cite{bib:flavianet}.} 
    \label{fig:kplusbr}
  \end{center}
\end{figure}

The Flavianet Kaon Working Group~\cite{bib:flaviawww} has performed
a global fit to all kaon branching fractions, lifetimes, and form factor slopes~\cite{bib:flavianet}.
In this fit, about 50 measurements of all published and preliminary results of most former and all recent experiments
were taken into account. The sum of branching fractions of each $K$ meson was constrained to be 1.
The fit results of the main $K_L$ and $K^\pm$ branching ratios are shown in 
Figs.~\ref{fig:klbr} and \ref{fig:kplusbr}. 
They are substantially more precise than the current PDG averages~\cite{bib:pdg}, and,
in case of the charged kaon, shifted with respect to the PDG values, mainly because of the new
KLOE measurement of Br$(K^\pm \to \pi^\pm \pi^0)$.

\begin{figure}[t]
  \begin{center}
    \includegraphics[width=0.5\linewidth]{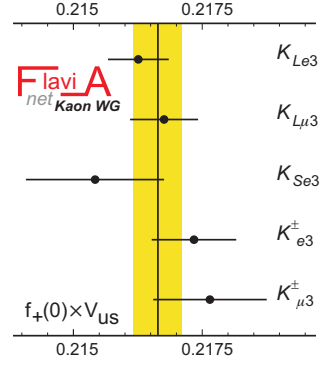}
    \caption{Measurements of $|V_{us}| \times f_+(0)$~\cite{bib:flavianet}.} 
    \label{fig:kaonbr}
  \end{center}
\end{figure}

The fit results for the semileptonic kaon branching fractions are given in the first column of
Table~\ref{tab:kaonbr}.
For determining $|V_{us}| \times f_+(0)$, the corrections $\delta^l_{SU(2)}$ and $\delta^l_{EM}$, which are of order $1\%$,
are taken from~\cite{bib:deltas}.
The final values of $|V_{us}| \times f_+(0)$, summarized in the second column of Table~\ref{tab:kaonbr},
agree well within the single decay channels (Fig.~\ref{fig:kaonbr}).
Averaging the results with correlations taken into account, gives
\be
|V_{us}| \times f_+(0) = 0.21661 \pm 0.00047,
\ee
where about equal parts of the uncertainty comes from experiment (BR's, lifetimes) and theory (mainly $\delta^l_{EM}$).

\begin{table}[t]
  \begin{center}
    \begin{tabular}{lllrccc}
                            &                      &                                    &                 & \multicolumn{3}{c}{\% error from} \\
                            & {\bf BR} [\%]        & $\mathbf{|V_{us}| \times f_+(0)}$  & \% err          & BR     & $\tau$ & $\Delta$ \\ \hline \hline
      $\mathbf{K_L e3}$     & $\mathbf{40.58(9)}$  & $\mathbf{0.21625(60)}$             & $\mathbf{0.28}$ & $0.09$ & $0.19$ & $0.15$   \\
      $\mathbf{K_L \mu3}$   & $\mathbf{27.06(6)}$  & $\mathbf{0.21675(66)}$             & $\mathbf{0.31}$ & $0.10$ & $0.18$ & $0.15$   \\
      $\mathbf{K_S e3}$     & $\mathbf{0.0705(9)}$ & $\mathbf{0.21542(134)}$            & $\mathbf{0.67}$ & $0.65$ & $0.03$ & $0.15$   \\
      $\mathbf{K^\pm e3}$   & $\mathbf{5.078(25)}$ & $\mathbf{0.21728(84)}$             & $\mathbf{0.39}$ & $0.26$ & $0.09$ & $0.26$   \\
      $\mathbf{K^\pm \mu3}$ & $\mathbf{3.365(27)}$ & $\mathbf{0.21758(111)}$            & $\mathbf{0.51}$ & $0.40$ & $0.09$ & $0.26$   \\ \hline \hline
      {\bf Average}$\!\!\!$ &                      & $\mathbf{0.21661(47)}$             &                 &        &        & 
    \end{tabular}
    \caption{World average branching fractions of semileptonic kaon decays and determination of $|V_{us}| \times f_+(0)$~\cite{bib:flavianet}.} 
    \label{tab:kaonbr}
  \end{center}
\end{table}

For the final determination of $\Vus$, the currently most precise value of 
$f_+(0) = 0.964 \pm 0.005$ from the UKQCD/RBC collaboration~\cite{bib:UKQCD} is taken. 
This yields
\be
|V_{us}| = 0.2246 \pm 0.0012.
\ee

Another possibility for the determination of $|V_{us}|$ is the comparison of the leptonic
charged kaon and pion decay widths. The KLOE collaboration has recently presented
a new precise measurement of Br$(K^\pm \to \mu^\pm \nu) = 0.6366 \pm 0.0017$~\cite{bib:KLOEKmu2}, which leads
to a new world average of~\cite{bib:flavianet}
\be
\Br(K^\pm \to \mu^\pm \nu) = 0.6357 \pm 0.0011.
\ee
Using $\tau_{K^\pm}$ from the global fit to all available kaon data~\cite{bib:flavianet}
and $\Gamma(\pi_{\mu2}) = 38.408(7)$~$\mu$s$^{-1}$~\cite{bib:pdg}, results
in $(|V_{us}|/|V_{ud}|) / (f_K/f_\pi) = 0.2760 \pm 0.0006$.
The currently best computation of the ratio of decay constants was performed by the MILC-HPQCD collaboration~\cite{bib:MILC},
giving $f_K/f_\pi = 1.189(7)$.
This results in 
\be
|V_{us}|/|V_{ud}| =  0.2321 \pm 0.0015,
\ee
with the error being dominated by $f_K/f_\pi$.

\begin{figure}[t]
  \begin{center}
    \includegraphics[width=0.7\linewidth]{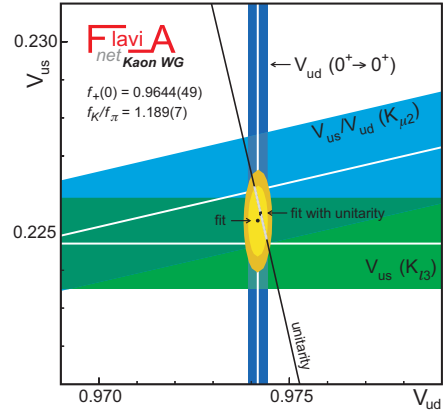}
    \caption{Measurements of $|V_{us}|$ versus $|V_{ud}|$~\cite{bib:flavianet}. 
             The black line is the SM prediction from unitarity of the CKM matrix.} 
    \label{fig:vusvud}
  \end{center}
\end{figure}

Figure~\ref{fig:vusvud}~\cite{bib:flavianet} shows the measurements of $\Vus$ and $\Vus/|V_{ud}|$ together with
the $|V_{ud}|$ value from $0^+ \to 0^+$ nuclear transitions and the SM prediction.
The data are in excellent agreement with CKM unitarity.

\begin{figure}[t]
  \begin{center}
    \includegraphics[width=0.7\linewidth]{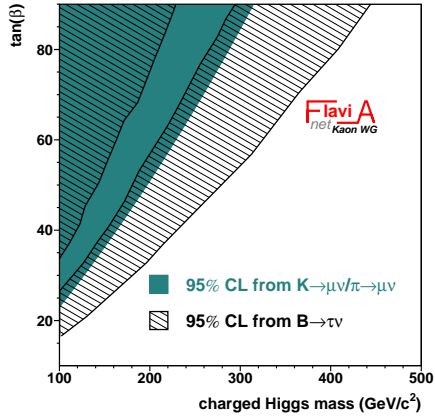}
    \caption{Exclusion limits on a SUSY charged Higgs mass and $\tan \beta$ from $K_{\mu2}$ decays~\cite{bib:flavianet}.
             Shown are also the corresponding limits from $B \to \tau \nu$ decays~\cite{bib:btaunu}.}
    \label{fig:higgsmass}
  \end{center}
\end{figure}

Finally, the precise measurement of Br$(K_{\mu2})$ can be used to set a 
stringent limit on the mass of a possible charged Higgs boson.
In the Standard Model, the ratio
\be
R_{l23} \equiv \left| \frac{V_{us}(K_{\mu2})}{V_{us}(K_{l3})} \times \frac{V_{ud}(0^+ \to 0^+)}{V_{ud}(\pi_{\mu2})} \right|
\ee
has to be equal to 1. In e.g.\ Super Symmetry, however, the leptonic kaon and pion
decay widths may be altered by exchange of a charged Higgs boson in addition to
$W$ exchange, yielding~\cite{bib:Kmu2higgs}
\be
R_{l23} = \left| 1 - \frac{m_{K^+}^2}{m_{H^+}^2} \right( 1 - \frac{m_{\pi^+}^2}{m_{K^+}^2} \left) \frac{\tan^2 \beta}{1 + \epsilon_0 \tan \beta} \right|.
\ee
From experimental data, using CKM unitarity for $K_{l3}$, and with $\frac{f_K}{f_\pi}/f_+(0)$ from the lattice~\cite{bib:flavianet},
\be
R_{l23} = 1.004 \pm 0.007,
\ee
in perfect agreement with the SM prediction.
The corresponding exclusion limits on the charged Higgs mass and $\tan \beta$ are competitive
to the limits from $B \to \tau \nu$ decays (Fig.~\ref{fig:higgsmass}).

\section{New Measurements of $\Gemu$}

The ratio $R_K = \Gemu$ can be precisely calculated within the Standard Model. Neglecting radiative
corrections, it is given by 
\be
R_K^{(0)} = \frac{m_e^2}{m_\mu^2} \: \frac{(m_K^2 - m_e^2)^2}{(m_K^2 - m_\mu^2)^2} = 2.569 \times 10^{-5},
\ee
and reflects the strong helicity suppression of the electronic channel.
Radiative corrections have been computed within the model of vector meson dominance~\cite{bib:rosell},
yielding a corrected ratio of
\begin{eqnarray}
R_K & = & R_K^{(0)} ( 1 + \delta R_K^{\text{rad.corr.}} ) \nonumber \\
    & = & 2.569 \times 10^{-5} \times ( 0.9622 \pm 0.0004 ) \nonumber \\
    & = & ( 2.477 \pm 0.001 ) \times 10^{-5}.
\end{eqnarray}

Because of the helicity suppression of $K_{e2}$ in the SM, the decay amplitude is a prominent candidate
for possible sizeable contributions from new physics beyond the SM. Moreover, when normalizing to $K_{\mu2}$ decays,
it is one of the few kaon decays for which the SM-rate is predicted with very high accuracy.
Any significant experimental deviation from the prediction would immediately be evidence for new physics.
However, this new physics would need to violate lepton universality to be visible in the ratio $K_{e2}/K_{\mu2}$.

Recently it has been pointed out, that in a SUSY framework sizeable violations of lepton universality can be expected
in $K_{l2}$ decays~\cite{bib:masiero}. At tree level, 
lepton flavour violating terms are forbidden in the MSSM. 
Loop diagrams, however, should induce lepton flavour violating Yukawa couplings as  $H^+ \to l \nu_\tau$ 
to the charged Higgs boson $H^+$.
Making use of this Yukawa coupling, the dominant non-SM contribution to $R_K$ modifies the ratio to
\be
R_K^{\text{LFV}} \approx R_K^{\text{SM}} \left[ 1 + \left( \frac{m_K^4}{M_{H^\pm}^4} \right) \left( \frac{m_\tau^2}{M_e^2} \right) |\Delta_{13}|^2 \tan^6 \beta \right].
\label{eqn:susy}
\ee
The lepton flavour violating term $\Delta_{13}$ should be of the order of $10^{-4}-10^{-3}$, as expected from neutrino mixing. 
For moderately large $\tan \beta$ and $M_{H^\pm}$, SUSY contributions may therefore enhance $R_K$ by up to a few percent.
Since the additional term in Eqn.~\ref{eqn:susy} goes with the forth power of the meson mass, no similar effect
is expected in $\pi_{l2}$ decays.

\begin{figure}[t]
  \begin{center}
    \includegraphics[width=0.8\linewidth]{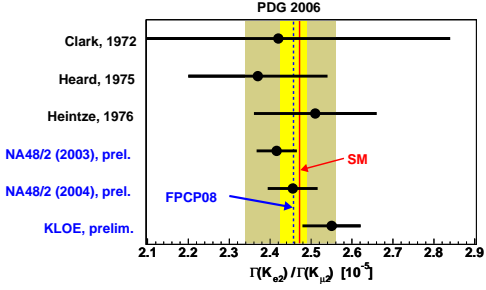}
    \caption{Results on $R_K = \Gemu$. Indicated are the current PDG average (brown),
             the average of all measurements (blue/yellow), and the SM prediction (red).}
    \label{fig:ke2kmu2}
  \end{center}
\end{figure}

Experimental knowledge of $K_{e2}/K_{\mu2}$ has been poor so far. The current world average
of $R_K = (2.45 \pm 0.11) \times 10^{-5}$ dates back to three experiments of the 1970s~\cite{bib:pdg}
and has a precision of less than 4\%.
However, now three new preliminary measurements have been reported by NA48/2 and KLOE (Tab.~\ref{tab:ke2kmu2}).
A preliminary result of NA48/2, based on about 4000 $K_{e2}$ events from the 2003 data set, 
was presented in 2005~\cite{bib:Ke2_2003}.
Another preliminary result, based on also about 4000 events, recorded in a minimum bias run period in 2004, 
was shown in 2007~\cite{bib:Ke2_2004}.
Both results have independent statistics and are also independent in the systematic uncertainties, 
as the systematics are either of statistical nature (as e.g.\ trigger efficiencies) or determined in
an independent way. 
Another preliminary result, based on about 8000 $K_{e2}$ events, was presented in 2007
by the KLOE collaboration~\cite{bib:Ke2_KLOE}.
Combining these new results with the current PDG value yields a current world average of
\be
R_K  = ( 2.457 \pm 0.032 ) \times 10^{-5},
\label{eqn:ke2kmu2}
\ee
in very good agreement with the SM expectation and, with a relative error of $1.3\%$,
a factor three more precise than the previous world average (Fig.~\ref{fig:ke2kmu2}).

\begin{table}[t]
  \begin{center}
    \begin{tabular}{lc}
      \hline \hline
                                                & $\Gemu$ $[10^{-5}]$  \\ \hline
      PDG 2006~\cite{bib:pdg}                   & $2.45 \pm 0.11$ \\
      NA48/2 prel.\ ('03)~\cite{bib:Ke2_2003}   & $2.416 \pm 0.043 \pm 0.024$ \\
      NA48/2 prel.\ ('04)~\cite{bib:Ke2_2004}   & $2.455 \pm 0.045 \pm 0.041$ \\
      KLOE prel.~\cite{bib:Ke2_KLOE}            & $2.55 \pm 0.05 \pm 0.05$ \\ \hline
      SM prediction                             & $2.472 \pm 0.001$ \\
      \hline \hline
    \end{tabular}
    \caption{Results and prediction for $R_K = \Gemu$.}
    \label{tab:ke2kmu2}
  \end{center}
\end{table}

\begin{figure}[t]
  \begin{center}
    \includegraphics[width=0.7\linewidth]{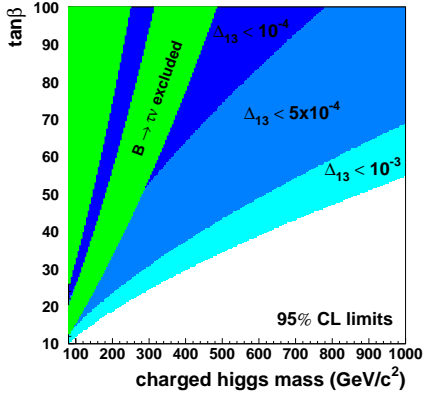}
    \caption{Exclusion limits at $95\%$ CL on $\tan \beta$ and the charged Higgs mass $M_{H^\pm}$ from $\Remu$ for different values of $\Delta_{13}$. 
             Values $M_{H^\pm} < 80$~GeV/$c^2$ are excluded by direct searches. 
             Indicated are also the exclusion limits from $B \to \tau \nu_\tau$ decays~\cite{bib:btaunu}.}
    \label{fig:susylimit}
  \end{center}
\end{figure}

In the SUSY framework discussed above, this result gives strong constraints for $\tan \beta$ and $M_{H^\pm}$ (Fig.~\ref{fig:susylimit}).
For a moderate value of $\Delta_{13} \approx 5 \times 10^{-4}$, $\tan \beta > 50$
is excluded for charged Higgs masses up to 1000~GeV/$c^2$ at 95\% CL.
These exclusion limits can be compared with SUSY limits obtained from $B \to \tau \nu$ decays.
The exclusion limits on $\tan \beta$ and $M_{H^\pm}$, obtained from
the current BELLE and BaBar average of $\Br(B \to \tau \nu) = (1.42 \pm 0.44) \times 10^{-4}$~\cite{bib:btaunu},
are superposed in Fig.~\ref{fig:susylimit}. 
In general, the limit obtained from $\Remu$ is stronger than those from $B \to \tau \nu_\tau$. However,
as the latter are lepton flavour conserving, they do not need an assumption on the value $\Delta_{13}$.

In the near future, further improvements in the knowledge of $R_K$ are expected.
The preliminary KLOE measurement~\cite{bib:Ke2_KLOE} 
is statistically dominated by MC statistics and has a conservative estimate of the systematic uncertainty.
Improving both these items, adding the remaining KLOE statistics, and also adding events with an 
additional reconstruction method, should reduce the overall relative uncertainty down to 1\%.

An improvement of a further factor of three is the goal of the NA62 collaboration at CERN. 
NA62, successor of NA48, plans to measure the very rare decay $\kppinunu$ in the mid-term future~\cite{bib:P326}.
In a first phase, about four months of data taking for the measurement of $\Remu$ took place in 2007,
with about 150000 $K_{e2}$ decays being recorded in total.
Preliminary results from this measurement are expected in the near future.

\section{Search for Direct $\CP$ Violation in $K^\pm$ Decays}

The violation of $CP$ symmetry has always been of great interest in particle physics, 
not only because of its cosmological implications, 
but also for its potential for the discovery of new physics beyond the Standard Model.
In particular, $\CP$ violation in decays (direct $\CP$ violation, as opposed to indirect $\CP$ violation and mixing induced $\CP$ violation) 
is sensitive to contributions of possible new physics.

For the decay of charged kaons into three pions, the obvious manifestation of direct $\CP$ violation would be an asymmetry in the partial rates.
However, measuring rate differences is experimentally almost impossible, 
as the total fluxes would have to be known to sub per mille precision.
Instead, experiments search for differences in the Dalitz plot slopes between $K^+$ and $K^-$ decays.

The $K^\pm \to 3\pi$ matrix element squared is conventionally parametrized as~\cite{bib:pdg} 
\be
|M(u,v)|^2 \sim 1 + gu + hu^2 + kv^2 + \cdots,
\ee
where $g$, $h$, and $k$ are the so-called linear and quadratic Dalitz plot slope parameters (with $|h|,|k| \ll |g|$),
and the two kinematic variables $u = (s_3-s_0)/m_{\pi^+}^2$ and $u = (s_2-s_1)/m_{\pi^+}^2$
with $s_i = (p_K - p_{\pi_i})^2$ and $s_0 = (s_1 + s_2 + s_3)/3$, and the index $i=3$ belonging to the ``odd''-charged pion.
Direct $\CP$ violation would manifest itself in a difference between the linear slope parameters
$g^+$ and $g^-$ for $K^+$ and $K^-$ decays, respectively, which is usually expressed in a corresponding
slope asymmetry
\be
A_g = \frac{g^+ - g^-}{g^+ + g^-} = \frac{\Delta g}{2 g},
\ee
where $g$ is the average linear slope.
The SM prediction for the size of $A_g$ is
of ${\cal O}(10^{-6})-{\cal O}(10^{-5})$~\cite{bib:gamiz}, while new scenarios beyond the SM may enhance
$\CP$ violation up to a few $10^{-4}$~\cite{bib:K3piNP}.

\begin{figure}[t]
  \begin{center}
    \includegraphics[width=0.52\linewidth]{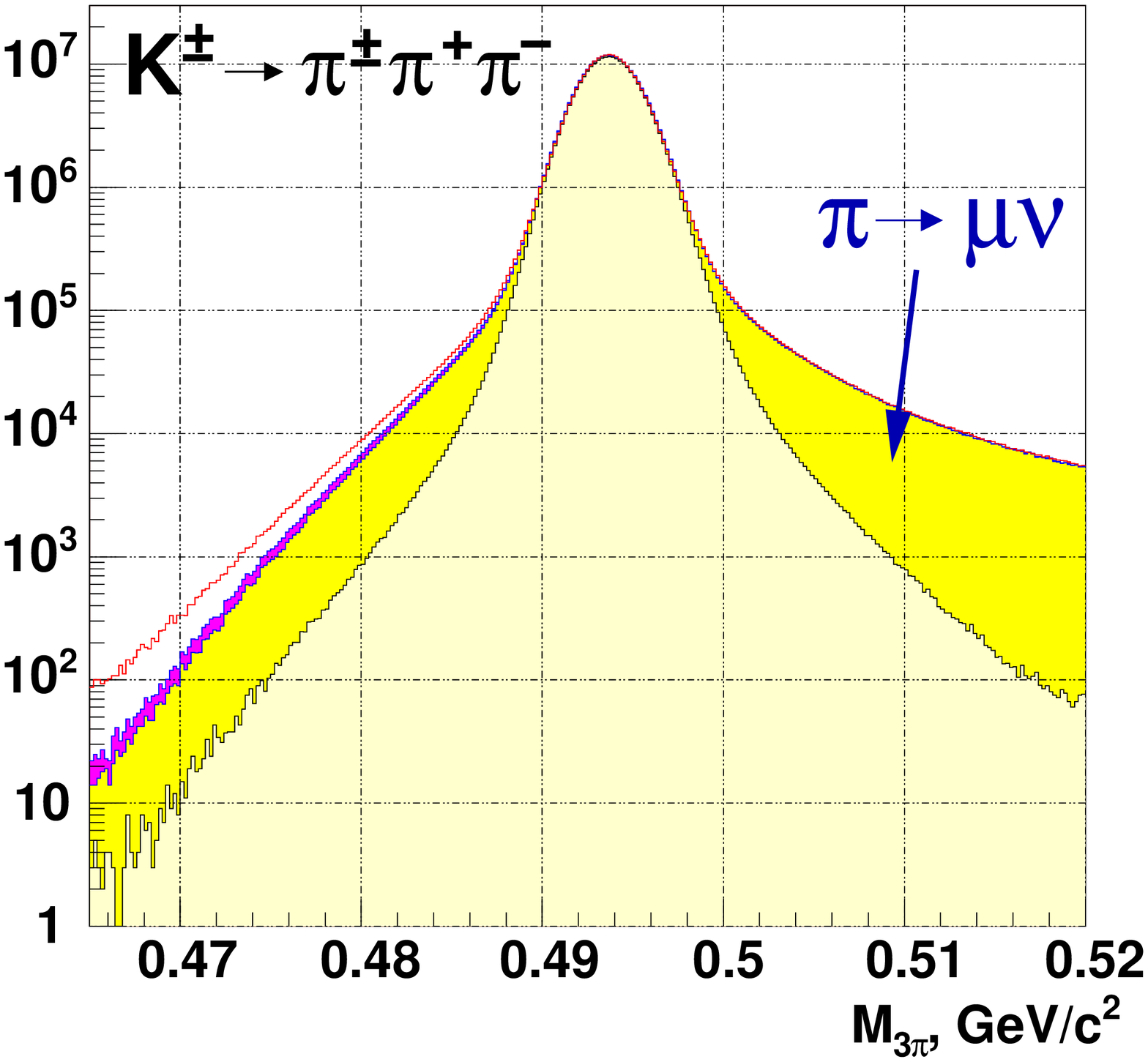}
    \includegraphics[width=0.465\linewidth]{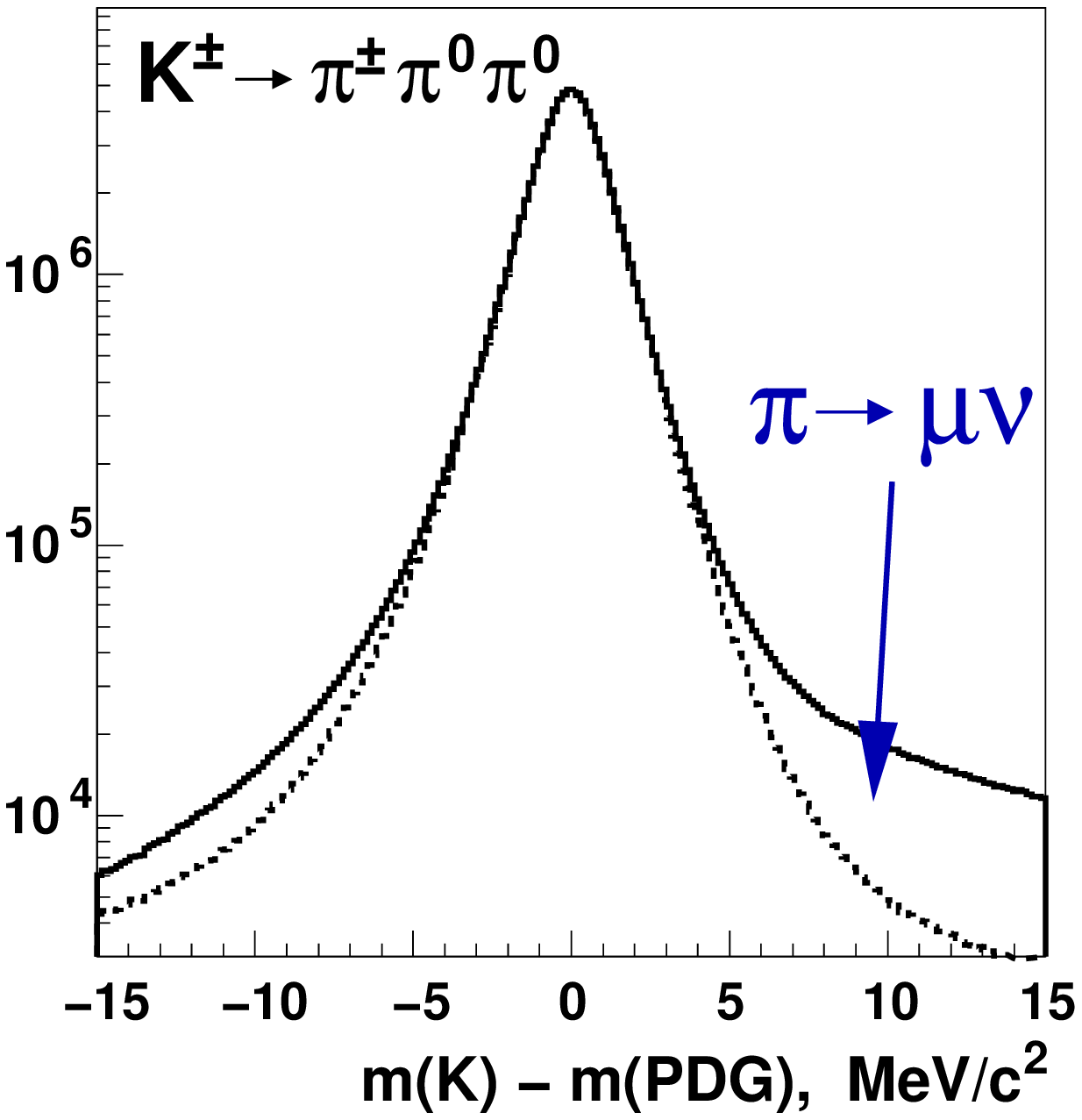}
    \caption{Reconstructed invariant $3\pi$ mass of $K^\pm \to \pi^\pm \pi^+ \pi^-$ (left)
             and $K^\pm \to \pi^\pm \pi^0 \pi^0$ (right) decays.}
    \label{fig:K3pimass}
  \end{center}
\end{figure}

\begin{figure}[t]
  \begin{center}
    \includegraphics[width=0.51\linewidth]{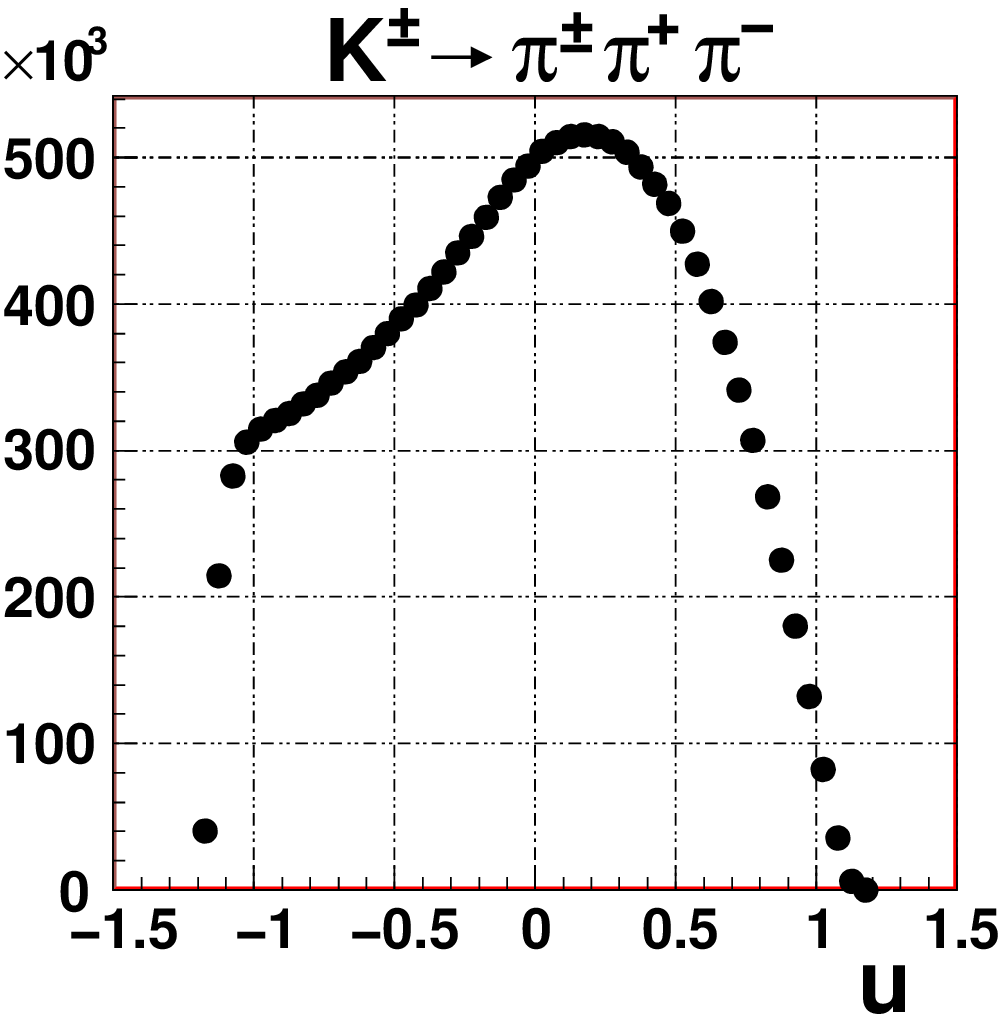}
    \includegraphics[width=0.475\linewidth]{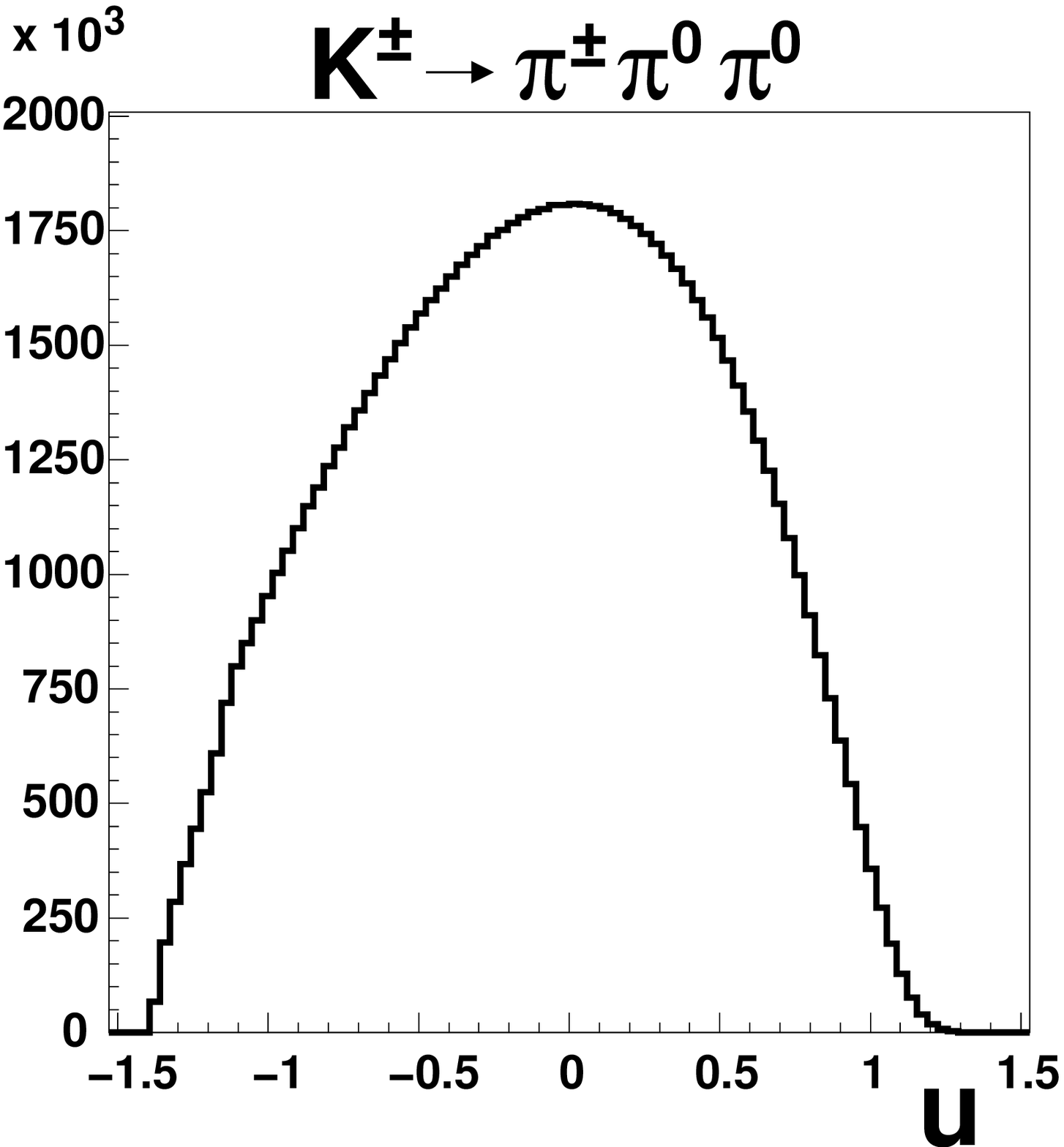}
    \caption{Distributions of the Dalitz plot parameter $u = (s_3-s_0)/m^2_{\pi}$ for $K^\pm \to \pi^\pm \pi^+ \pi^-$ (left)
             and $K^\pm \to \pi^\pm \pi^0 \pi^0$ (right) decays.}
    \label{fig:K3piu}
  \end{center}
\end{figure}

The NA48/2 experiment was specifically designed to measure the Dalitz plot 
asymmetries between $K^+$ and $K^-$ decays~\cite{bib:NA48asym}. $K^+$ and $K^-$ beams
were produced together by proton collisions on a single beryllium target 
and selected by an achromat system to have momenta of $p_{K^\pm} = (60 \pm 3)$~GeV/$c$.
Superimposed on each other, the two beams entered the decay region, where $K^+$ and $K^-$ decays were recorded simultaneously.
To further symmetrize the set-up, regular changes of the beam-line achromat and the detector spectrometer 
polarities were performed during the data taking periods.
In total, $3.1 \times 10^9$ $K^\pm \to \pi^\pm \pi^+ \pi^-$ events and 
$9.1 \times 10^7$ $K^\pm \to \pi^\pm \pi^0 \pi^0$ events were reconstructed from the final NA48/2 data set.
Both decays were selected practically background-free.
The invariant $3\pi$ mass distributions are shown in Fig.~\ref{fig:K3pimass}. 
The corresponding distributions of the Dalitz plot variable $u$ are shown in Fig.~\ref{fig:K3piu}.

\begin{figure}[t]
  \begin{center}
    \includegraphics[width=0.7\linewidth]{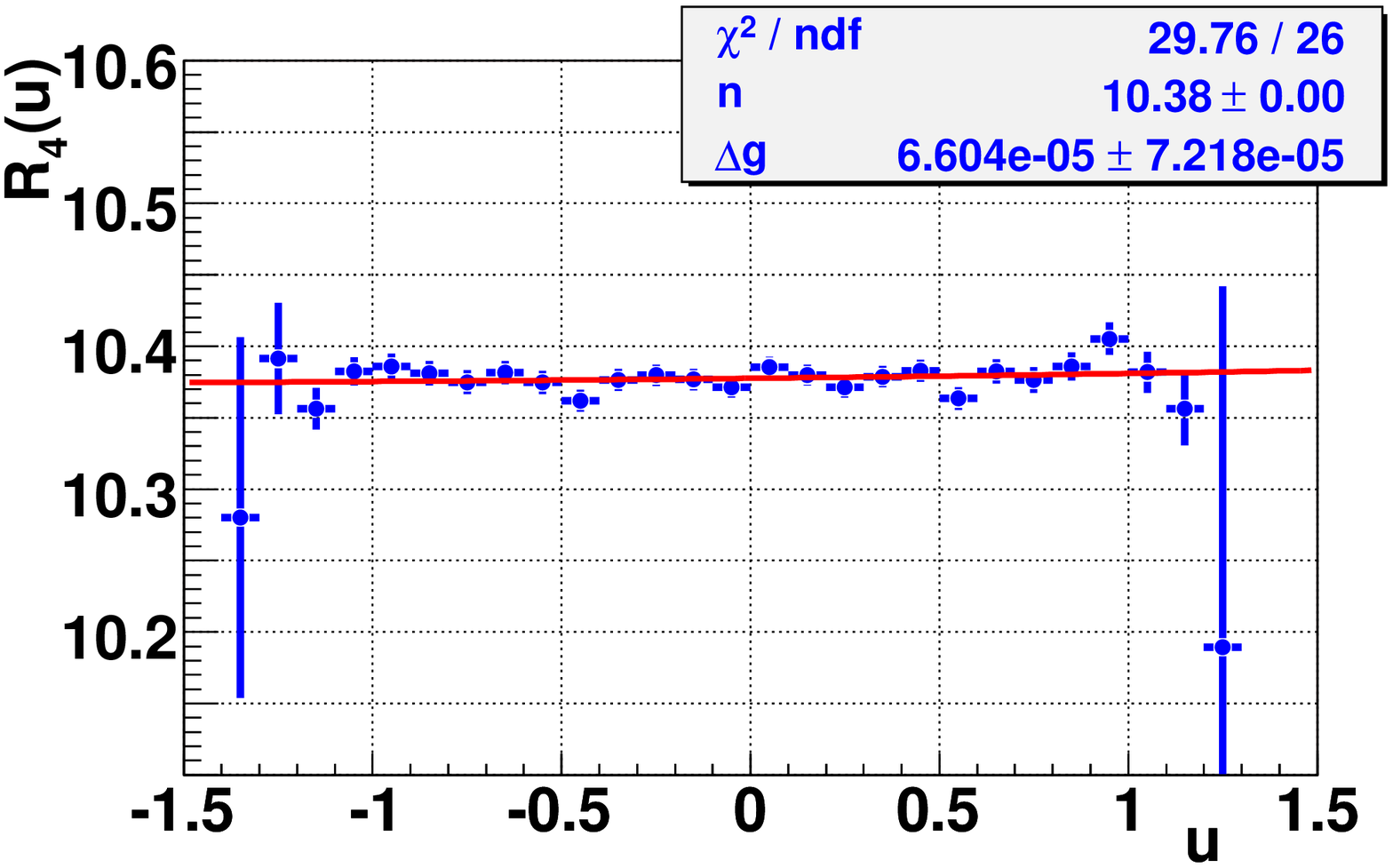} \\*[2mm]
    \includegraphics[width=0.7\linewidth]{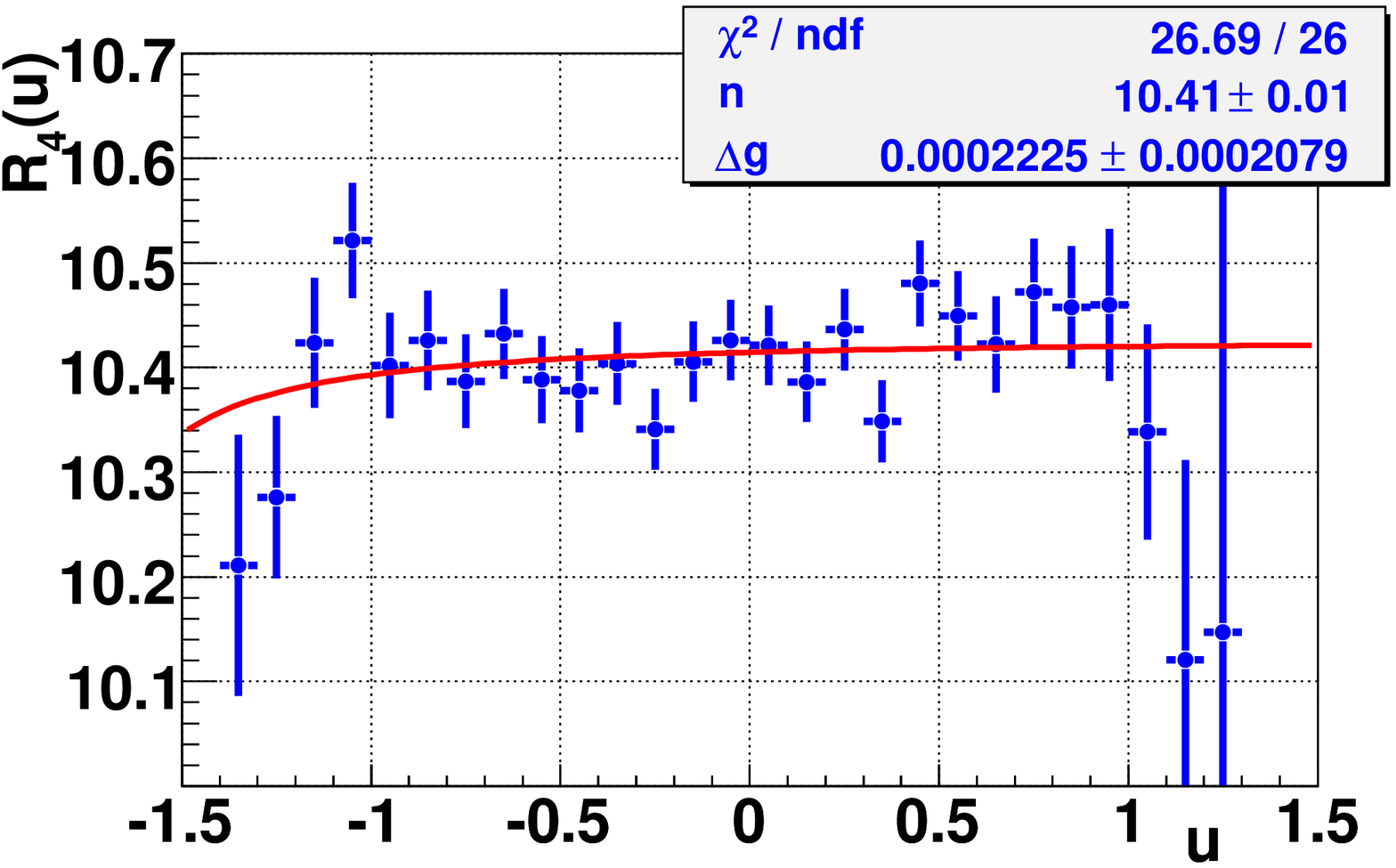}
    \caption{Distributions of the quadruple ratio $R_4(u)$ for $K^\pm \to \pi^\pm \pi^+ \pi^-$ (top)
             and $K^\pm \to \pi^\pm \pi^0 \pi^0$ (bottom) data together with the best fit.}
    \label{fig:K3pifit}
  \end{center}
\end{figure}

To cancel possible systematic biases due to achromat and spectrometer polarity reversals, four single ratios
$R_{UJ}(u)$, $R_{US}(u)$, $R_{DJ}(u)$, and $R_{DS}(u)$ 
between the $K^+$ and $K^-$ $u$-spectra were built, where in each ratio only data were considered 
with both kaons either going through the upper (U) or lower (D) achromat path and
being deflected by the spectrometer either to the left (Jura, J) or to the right (Sal\`eve, S) side.
Finally, a Òquadruple ratioÓ 
\begin{eqnarray}
R_4(u) & = & R_{UJ}(u) \cdot R_{US}(u) \cdot R_{DJ}(u) \cdot R_{DS}(u)  \nonumber \\*[1mm]
       & \propto & \left( \frac{\Delta g \, u}{1 + g \, u + h \, u^2} \right)^4
\end{eqnarray}
was evaluated in the fit. In this ratio, in contrast to the single ratios, also possible time variation of run and detector conditions cancel.
The $R_4(u)$ distributions for both modes are shown in Fig.~\ref{fig:K3pifit}. The fit resulted in asymmetries of~\cite{bib:NA48asym}
\begin{eqnarray}
A_g^{\pi^\pm \pi^+ \pi^-} \! \! & = & (-1.5 \pm 1.5_{\text{stat}} \pm 0.9_{\text{trig}} \pm 1.3_{\text{syst}}) \cdot 10^{-3}, \nonumber \\
A_g^{\pi^\pm \pi^0 \pi^0} & = & ( 1.8 \pm 1.7_{\text{stat}} \pm 0.6_{\text{syst}}) \cdot 10^{-3},
\end{eqnarray}
where the trigger uncertainty is of purely statistical nature.
For both results, the statistical uncertainties dominate.
They are an order of magnitude more precise than previous measurements and give no indication of
possible non-SM $\CP$-violating amplitudes in these channels.

\bigskip 

\begin{thebibliography}{99} 

\bibitem{bib:Kl3formula} E.~Blucher {\it et al.},
                         arXiv:hep-ph/0512039 (2005). 

\bibitem{bib:flaviawww}  {\tt http://www.lnfn.infn.it/wg/vus.}

\bibitem{bib:KLOEbr}   F.~Ambrosino {\it et al.} (KLOE Collaboration),
                       \Journal{\JHEP}{0802}{098}{2008}, and references therein.
                   
\bibitem{bib:NA48br}   J.R.~Batley {\it et al.} (NA48/2 Collaboration),
                       \Journal{\EPJ}{50}{329}{2007}; 
                       Erratum \Journal{\em ibid.}{52}{1021}{2007}.
                   
\bibitem{bib:ISTRAbr}  V.I.~Romanovsky {\it et al.} (ISTRA$+$ Collaboration),
                       arXiv:0704.2052 [hep-ex] (2007). 

\bibitem{bib:flavianet} M.~Antonelli {\it et al.} (Flavianet Kaon Working Group),
                       arXiv:0801.1910 [hep-ph] (2008). 
                   
\bibitem{bib:pdg}      W.-M.~Yao {\it et al.} (Particle Data Group),
                       \Journal{\JPG}{33}{1}{2006}.

\bibitem{bib:deltas}   V.~Cirigliano {\it et al.}, 
                       \Journal{\EPJ}{35}{53}{2004};
                       \Journal{\EPJ}{23}{121}{2002};
                       T.C.~Andre, 
                       \Journal{\em Annals~Phys.}{322}{2518}{2007};
                       V.~Cirigliano, M.~Giannotti, and H.~Neufeld,
                       work in preparation. 

\bibitem{bib:UKQCD}    D.J.~Antonio {\it et al.} (RBC and UKQCD Collaborations),
                       arXiv:hep-lat/0702026 (2007)

\bibitem{bib:KLOEKmu2} F.~Ambrosino {\it et al.} (KLOE Collaboration),
                       \Journal{\PLB}{632}{76}{2006}.
                   
\bibitem{bib:MILC}     E.~Follana, C.T.H.~Davies, G.P.~Lepage, and J.~Shigemitsu  (HPQCD Collaboration),
                       \Journal{\PRL}{100}{062002}{2008}.

\bibitem{bib:Kmu2higgs} G.~Isidori and P.~Paradisi,
                        \Journal{\PLB}{639}{499}{2006};
                        W.S.~Hou,
                        \Journal{\PRD}{48}{2342}{1993};
                        A.G.~Akeroyd and S.~Recksiegel,
                        \Journal{\JPG}{29}{2311}{2003}.

\bibitem{bib:btaunu}   K.~Ikado {\it et al.} (BELLE Collaboration), 
                       \Journal{\PRL}{97}{251802}{2006};
                       B.~Aubert {\it et al.} (BABAR Collaboration),
                       \Journal{\PRD}{76}{052002}{2007};

\bibitem{bib:rosell}   V.~Cirigliano and I.~Rosell,
                       \Journal{\PRL}{99}{231801}{2007}.

\bibitem{bib:masiero}  A.~Masiero, P.~Paradisi, and R.~Petronzio,
                       \Journal{\PRD}{74}{011701(R)}{2006}.

\bibitem{bib:Ke2_2003} L.~Fiorini, 
                       \Journal{\PoS}{HEP2005}{288}{2006};
                       L.~Fiorini, ph.D.~thesis, Pisa (2005);
                       
                        
\bibitem{bib:Ke2_2004} V.~Kozhuharov, 
                       \Journal{\PoS}{KAON}{049}{2007};

\bibitem{bib:Ke2_KLOE} F.~Ambrosino {\it et al.} (KLOE Collaboration), 
                       \Journal{\PoS}{KAON}{050}{2007};

\bibitem{bib:P326}     G.~Anelli {\it et al.} (P326 Collaboration),
                       CERN-SPSC-2005-013 (2005).

\bibitem{bib:gamiz}    E.~G\'amiz, J.~Prades, and I.~Scimeni,
                       \Journal{\JHEP}{10}{042}{2003};
                       G.~F\"aldt, E.~Shabalin,
                       \Journal{\PLB}{635}{295}{2006};

\bibitem{bib:K3piNP}   E.P.~Shabalin,
                       ITEP preprint {\bf 8-98} (1998);
                       G.~D'Ambrosio, G.~Isidori, and G.~Martinelli,
                       \Journal{\PLB}{480}{164}{2000}. 

\bibitem{bib:NA48asym} J.R.~Batley {\it et al.} (NA48/2 Collaboration),
                       \Journal{\EPJ}{52}{875}{2007}.
                   


\end{thebibliography}

\end{document}